\documentclass[%
 reprint,
superscriptaddress,
showpacs,preprintnumbers,
 amsmath,amssymb,
 aps,
prb,
]{revtex4-1}

\usepackage{graphicx}
\usepackage{dcolumn}
\usepackage{bm}

\begin{document}

\preprint{APS/123-QED}

\title{Spin force and the generation of sustained spin current in time-dependent Rashba and Dresselhauss systems}

\author{Cong Son Ho}
 \affiliation{
Department of Electrical and Computer Engineering, National University of Singapore,4 Engineering Drive 3, Singapore 117576, Singapore.
}
 \affiliation{
Data Storage Institute, Agency for Science, Technology and Research (A*STAR),
DSI Building, 5 Engineering Drive 1,Singapore 117608, Singapore.
}
\author{Mansoor B. A. Jalil}%
\affiliation{
Department of Electrical and Computer Engineering, National University of Singapore,4 Engineering Drive 3, Singapore 117576, Singapore.
}
\author{Seng Ghee Tan(陳繩義)}
\affiliation{
Department of Electrical and Computer Engineering, National University of Singapore,4 Engineering Drive 3, Singapore 117576, Singapore.
}
\affiliation{
Data Storage Institute, Agency for Science, Technology and Research (A*STAR),
DSI Building, 5 Engineering Drive 1,Singapore 117608, Singapore.
}

\date{\today}

\begin{abstract}
The generation of spin current and spin polarization in a 2DEG structure is studied in the presence of Dresselhaus and Rashba spin-orbit couplings (SOC), the strength of the latter being modulated in time by an ac gate voltage. By means of the non-Abelian gauge field approach, we established the relation between the Lorentz spin force and the spin current in the SOC system, and showed that the longitudinal component of the spin force induces a transverse spin current. For a constant (time-invariant) Rashba system, we recover the universal spin Hall conductivity of $\frac e{8\pi}$, derived previously via the Berry phase and semiclassical methods. In the case of a time-dependent SOC system, the spin current is sustained even under strong impurity scattering. We evaluated the ac spin current generated by a time-modulated Rashba SOC in the absence of any dc electric field. The magnitude of the spin current reaches a maximum when the modulation frequency matches the Larmor frequency of the electrons.

\end{abstract}

\maketitle


\section{INTRODUCTION}
The generation of a sustained spin polarized current constitutes a key requirement in practical spintronics. In two-dimensional electron gas (2DEG) systems with Rashba spin-orbit coupling (SOC), spin precession about the Rashba field and the accompanying Zitterbewegung-like motion would usually result in zero net transverse spin current on average.\cite{Shen05} However, a net spin polarization can be generated in the transverse direction via the spin Hall effect. Based on the time-gauge field \cite{Fu10} or the semi-classical geometric force \cite{Ah92} descriptions, it can be shown that a vertical (out-of-plane) spin polarization can be sustained under adiabatic conditions, albeit under ballistic limit only. Other methods of generating spin current which have been proposed include the application of a non-uniform external magnetic field in a SOC medium \cite{Tan08} and the use of optical excitation.\cite{Sun03,Sherman05} Recently, an all-electrical method has been also investigated by utilizing a time-varying gate voltage.\cite{Mal03,Tang05,Liang09,Zhang13} The idea is based on the fact that the Rashba SOC can be tuned by changing the applied gate voltage.\cite{Datta90,Nit97,Gru00} However, it is pointed out that, in the Rashba system, only the in-plane polarized spin currents are generated.\cite{Mal03,Zhang13} Therefore, it would be interesting if the out-of-plane polarized component of the spin current can be generated by applying this method. 

In this article, we will present a theoretical study of the spin current generation in 2DEG system that combines both time-dependent Rashba and Dresselhauss SOCs. In frame work of non-Abelian gauge field, we will point out that  time-modulated gate voltage can indeed induce a spin current polarized in the out-of-plane direction of the 2DEG. This spin current can be maintained even under non-ballistic situation, i.e., in the presence of impurities. Moreover, we will show that such combined SOC systems can be utilised as spin pumping sources to generate spin current without applying bias.

\section{SPIN FORCE AND SPIN HALL CURRENT}
\begin{figure}
 \includegraphics[width=0.3\textwidth]{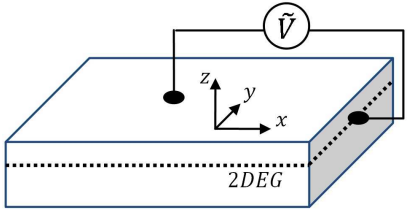}
 \caption{Schematic diagram of a 2DEG system with Rashba and Dresselhaus SOC. The strength of the Rashba SOC can be modulated in time by applying an ac gate bias.\label{Fig.1.}}
 \end{figure}

We first consider a system with Dresselhaus and Rashba SOC, with the Rashba coupling being time-dependent (shown schematically in Fig. \ref{Fig.1.}). The Hamiltonian of the system is:
\begin{equation} \label{eq1}
H=\frac{p^2}{2m}+H_R(t)+ H_D + V({\bf r}),
\end{equation}
where $m\ $is the effective electron mass, $H_D=\beta\left(p_y\sigma_y-p_x\sigma_x\right)$ is the Dresselhaus SOC, and $H_R(t)= \alpha \left(t\right)\left(p_x{\sigma }_y-p_y{\sigma }_x\right)$ is the Rashba SOC, with the Rashba coupling consisting of static and time-varying components, i.e., $\alpha \left(t\right)={\alpha }_0+{\alpha }_1(t)$. The last term in Eq. \eqref{eq1} is the potential energy which may include the external applied electric field and the impurity field.

To the first order in SOC couplings, Eq. \eqref{eq1} can be rewritten in the form of the Yang-Mills Hamiltonian $H_{YM}=\frac{1}{2m}{\left({\bf p}-e{\bf {\cal A}}\right)}^2$, where $\bf{{\cal A}}$ is the gauge field given by
\begin{equation} \label{eq2}
{\mathbf {\cal A}}=\left({\cal A}_x,{\cal A}_y,0\right)=\frac{m}{e}\left(-\alpha{\sigma }_y+\beta\sigma_x,\alpha{\sigma }_x-\beta\sigma_{y},0\right),
\end{equation}
and the corresponding curvature is
\begin{equation} \label{eq3}
{\cal F}_{\mu \nu }={\partial }_{\mu }{\cal A}_{\nu }-{\partial }_{\nu }{\cal A}_{\mu }-\frac{ie}{\hbar }\left[{\cal A}_{\mu },{\cal A}_{\nu }\right].
\end{equation}
The physical fields extracted from the curvature tensor can be considered as the effective magnetic and electric fields, i.e.,
\begin{eqnarray}
{\cal B}={\cal F}_{xy}\hat z=\frac{{2m}^2{(\alpha }^2-\beta^2)\ }{e\hbar }{\sigma }_z {\hat z} ,\label{eq4}  \\
{\cal E}=\partial_t{\cal A}=-\frac{m}{e}{\dot \alpha } ({\bf\sigma}\times{\hat z}),\label {eq4b}
\end{eqnarray}
where ${\dot \alpha }={\partial\alpha }/{\partial t}$, $\hat z$ is the unit vector in the $z$-direction. These Yang-Mills fields exert a spin-dependent force on the electron, with the Lorentz-like magnetic field acting in the direction transverse to the electron motion as ${\bf F}_{YM}=e({\cal E}+{\bf v}\times {\bf\cal B})$, where $\bf v$ is the velocity.  Note that the Lorentz force term is proportional to the spin current polarized along the $z$-direction, which is given  by $\ {\bf j}^z_{s}=({\hbar }/{4})\{{\bf v},{\sigma }_z\}$. Therefore,  one can arrive at the expression for the quantum mechanical force:
\begin{equation} \label{eq7}
{\bf F}=\frac{4m^2\left({\alpha }^2-\beta^2\right)}{{\hbar }^2}({\bf j}^z_s\times {\hat z})-m{\dot \alpha\ }({\bf\sigma}\times{\hat z})-{\nabla V}.
\end{equation}
From the above, we see the interplay between spin force, spin current and the time modulation of the Rashba SOC. The spin current is dependent on either the spin force or the time-modulated RSOC, and would still be present if either one of the two terms vanishes.  One can directly evaluate the spin current from its definition by solving the spin dynamics of the system.\cite{Sino04,Shen04} However, this method may be quite involved if the Hamiltonian is non-trivially time-dependent. Alternatively, the spin current may be obtained by evaluating the force by some other means, and substituting its expression into Eq. \eqref{eq7}. In the following, we evaluate the force by classical consideration, and subsequently apply the force expression to determine the spin current.

We consider the instantaneous eigenstate and eigen-energy of the time-dependent SOC Hamiltonian $H_{SOC}|\psi_n(t)\rangle=E_n(t)|\psi_n(t)\rangle$,  given by
\begin{eqnarray}
\left.{|\psi }_{\pm }\right\rangle &=&\frac{1}{\sqrt{2}}\left(\genfrac{}{}{0pt}{}{1}{{\pm \ ie}^{i\gamma}}\right),\\
E_{\pm }&=&\frac{p^2}{2m}\mp \Delta E ,
\end{eqnarray}
in which ${\tan \gamma=\left(\alpha p_y+\beta p_x\right)/\left(\alpha p_x+\beta p_y\right)\ }$,  $\Delta {E}=p({\alpha}^2+{\beta }^2+2\alpha\beta {\sin  2\theta \ })^{1/2}$ with $\theta ={{\tan }^{{\rm -}{\rm 1}} p_y/p_x\ }$. The expectation value of an operator at time $t$ is defined as $\langle\hat{O}\rangle=\sum_n{\langle\psi_n|\hat{O}|\psi_i\rangle f_n}$, with $f_n=f(E_n)$ is the Fermi distribution function.

The velocity operator can be formally derived from the Hamiltonian as ${\bf v}=\partial H/\partial{\bf  p}$, and the force is ${\bf F}=d{\bf v}/dt$. At some instant $t$, the average force can be decomposed into two parts as
\begin{equation}\label{eq7c}
\langle {\bf F}\rangle={\bf F}^0+{\bf F}^t
\end{equation}
in which
\begin{eqnarray}
 {\bf F}^0=m&&\int\sum_{n=\pm} {\frac{d {\bf v}_n}{dt}  f_n},\label{eq7d}\\
 {\bf F}^t=-\frac{m}{2}\int {\dot\gamma}&&{\nabla_{{\bf p}}\gamma}  (E_+-E_-)(f_+-f_-) ,\label{eq7e}
\end{eqnarray}
where, ${\bf v}_n=\frac{\partial E_n({\bf p})}{\partial {\bf p}}$ is the velocity of electron belonging to the $n$th branch, and the integral is taken over momentum space. The first force term ${\bf F}^0$ is related to the evolution of the velocity over time due to the applied electric field and the modulating Rashba coupling, while the second component ${\bf F}^t$ is related to the explicit time-dependence of the eigen-state and will vanish if the Hamiltonian is time-independent.

{\it Constant Rashba coupling}-- We first consider the static case when $\alpha ={\alpha }_0$. The second term on the right hand side of Eq. \eqref{eq7} vanishes, hence the spin Hall current (SHC) is just dependent on the total force acting on the electrons. Since the eigen-energy and eigen-state are time-independent, its force component ${\bf F}^t$ vanishes. For simplicity, we consider an electric field applied along the $x$-direction, following which the longitudinal force in Eq.\eqref{eq7d} is readilyf found to be
\begin{equation} \label{eq11}
F_x=\left({\rm 1}\pm \frac{m{{\rm (}{\alpha_0 }^2{\rm -}{\beta }^{{\rm 2}}{\rm )}}^2{{\sin  \theta \ }}^2}{p{{\rm (}{\alpha_0 }^2{\rm +}{\beta }^{{\rm 2}}{\rm +}2\alpha_0 \beta {\sin  {\rm 2}\theta \ }{\rm )}}^{3{\rm /2}}}\right)\frac{{dp}_x}{dt},
\end{equation}
where $dp_x/dt=e{E}_x$. Comparing the above force expression with that obtained by the gauge formalism, i.e. Eq. \eqref{eq7}, we arrive at the formula of the spin current for the two eigenbranches:
\begin{equation} \label{eq12}
j^z_{s,y}\left({\mathbf p}\right)=\pm \frac{{e{E}}_x\left({\alpha_0}^2-{\beta }^2\right){\hbar }^2{\sin  {\theta }^2\ }}{4pm{({\alpha_0}^2+{\beta }^2+2\alpha_0\beta {\sin  2\theta \ })}^{{3}/{2}}}.
\end{equation}
The total SHC at zero temperature is then found by integrating over all occupied states in the momentum space:
\begin{eqnarray} \label{eq13}
J^z_{s,y}=\frac{1}{{(2\pi\hbar)}^2}\int^{p_+}_{p_-}{p dp}\int^{2\pi }_0{d\theta}\ j^z_{s,y}\left({\mathbf p}\right)
\end{eqnarray}
where, $p_+$ and $p_-$ are the Fermi momenta corresponding to the two eigenbranches, and the difference between them is given by:
$p_+-p_-=2m{\left({\alpha_0}^2+{\beta }^2+2\alpha_0\beta \ {\rm sin}2\theta \right)^{1/2}}. $
With this, the SHC expression simplifies to
\begin{equation} \label{eq15}
J^z_{s,y}=\frac{{\alpha_0}^2-{\beta }^2}{{|\alpha }^2_0-{\beta }^2|}\left(\frac{e}{8\pi}\right){E}_{x}.
\end{equation}
The above result is consistent with the well-known universal spin Hall conductivity of $\left({e}/{8\pi}\right)$ obtained previously by Sinova \textit{et al.},\cite{Sino04} and by Shen,\cite{Shen04} the latter by considering linear response (Kubo) transport theory and the Berry phase of the system. Up to this point, we have explicitly shown the relation between the spin Hall current and the spin force in a time-independent system. In Ref. \cite{Shen05}, the spin force concept is introduced to describe the spin dynamics in a Rashba SOC system. Specifically, the study focused on the \emph{transverse} component of the spin force and its effect on the Zitterbewegung dynamics of the electrons in the system. However, the connection between this spin force and the well-known spin Hall current was not established. In the above analysis, we have shown that it is the \emph{longitudinal} component of the spin force, whose magnitude is proportional to the driving electric field, which gives rise to the spin Hall current with the universal conductivity of $(e/8\pi)$. The longitudinal spin force is related to the Lorentz force of the Yang-Mills (non-Abelian) effective fields arising from the Rashba and Dresselhaus SOC effects. Our physical picture of the spin Hall effect is consistent with previous descriptions of the spin Hall effect based on different theoretical models.\cite{Sino04,Shen04,Fu10}
\section{EFFECT OF IMPURITY SCATTERING}
We now consider the effect on the spin current of impurities, which can be modeled as some randomly placed delta potentials: $V_{im}({\mathbf r})=V\sum^N_{i=1}{\delta\left({\mathbf r}-{{\mathbf R}}_i\right)}$. As has been pointed out previously,\cite{Inoue04} the spin Hall current would be suppressed in an infinitely large disordered system. In the steady state, the total force is zero, i.e., ${\bf F}=0$. Additionally, the external electric field which is the driving force for the spin current, is also effectively canceled by the effect of impurities.\cite{Ada05} Thus from Eq. \eqref{eq7}, the spin current vanishes if the Rashba coupling is constant in time, as expected.   However, in the presence of a time-modulated Rashba SOC strength, the $\dot{\alpha}$ term in Eq. \eqref{eq7} is non-zero, and this leads to a spin current of
\begin{equation} \label{eq16}
J^z_{s,i}=\frac{{\hbar }^2}{4m} \frac{{\dot \alpha } S_i}{\left({\alpha }^2-\beta^2\right)}  ,
\end{equation}
in which $S_i=\langle {\sigma }_i\rangle$ is the spin density, with $\langle \dots\rangle $ denoting the expectation values taken over all momentum and spin spaces, and impurity configurations. We see that the transverse spin current in any one direction depends on the spin polarization along that direction; in general the latter may be finite and hence the spin current can be sustained. On the other hand, the spin current can also be interpreted as the response to the effective electric field which is given by Eq. \eqref{eq4b}. Thus, we have $J^z_{s,i}={\sigma }_{ij}{\cal E}_j$, with the ``spin conductivity'' of ${\sigma }_{xy}=-{\sigma }_{yx}=\frac{e{\hbar }^2}{4m^2\left({\alpha}^2-{\beta}^2\right)}$. We consider a special case in which the Rashba and Dresselhaus coupling strengths are approximately the same, but with the former having an additional (small) time-dependent variation, i.e., the Rashba coupling is given by $\alpha(t)=\beta + \alpha_1 e^{i \omega t}$ with  $\alpha_1\ll\beta$. Then to the first order in $\alpha_1$, we can express the spin current in Eq. \eqref{eq16} as:
\begin{equation} \label{eq16a}
J^z_{s,i}(\omega) =\frac{{\omega\hbar}^2}{8m\beta} S_i(\omega).
\end{equation}
Thus, the spin current is approximately proportional to the modulation frequency, but is independent of the modulating amplitude $\alpha_1$.

\section{GATE-MODULATION INDUCED SPIN CURRENTS}
 Normally, when an electron with its spin aligned, say, vertically, is passed through the Rashba (Dresselhauss) 2DEG system, the spin will rotate about the in-plane Rashba field.\cite{Rashba84,Lommer88} This spin precession is associated with Zitterbewegung-like motion of the electron.\cite{Shen05,Biswas12} Under the influence of an electric field, the precessional dynamics is modified giving rise to a net spin polarization and hence a spin Hall current. Since the spin precession motion originates from the SOC interaction, this suggests that we induce a net spin polarization by modulating the SOC to control the spin precession without the need to apply any electric field. Thus, we will study the possibility to generate a sustained spin current under the influence of time-modulated SOC but in the absence of any dc electric field.

First, we separate the Hamiltonian in Eq. \eqref{eq1} into two parts: time-independent Hamiltonian $H_0=\frac{p^2}{2m}+{\alpha}_0\left(p_x{\sigma }_y-p_y{\sigma }_x\right)+\beta\left(p_y\sigma_y-p_x\sigma_x\right)$, and time-dependent Hamiltonian ${H_1\left(t\right)=\alpha }_1(t)\left(p_x{\sigma }_y-p_y{\sigma }_x\right)+e{\bf E.r}$. The evolution of electron spin state can be found by means of the first-order time-dependent perturbation theory, in which $H_1(t)$ is treated as the perturbation. We assume that the modulation is started at $t=0$; meanwhile the applied electric field is switched on adiabatically from the past as ${\bf E}=\lim_{s\rightarrow 0^+} {\bf E} e^{st}$.\cite{KL57} If the initial electron spin state is represented by the spinor $\chi(0)=c_+\left.{|\psi }_+\right\rangle +c_-\left.{|\psi }_-\right\rangle$, then its subsequent evolution up to time $t$ under the full Hamiltonian is given by
\begin{equation}
\left.|\chi (t)\right\rangle =\sum_{i=\pm }{c_i(t)e^{-\frac{i}{\hbar }E_it}\left.{|\psi }_i\right\rangle },
\end{equation}
in which $\left.{|\psi }_\pm\right\rangle$ and $E_\pm$ are eigen-states and eigen-energies of the unperturbed Hamiltonian. To the first order in both the dynamic Rashba component $\alpha_1$ and the electric field, the time-dependent coefficients $c_i(t)$ satisfy the following differential equations:
\begin{equation} \label{eq20}
i\hbar \frac{dc_i}{dt}=\sum_{j=\pm }{\left\langle {\psi }_i\mathrel{\left|\vphantom{{\psi }_i H_1\left(t\right) {\psi }_j}\right.\kern-\nulldelimiterspace}H_1\left(t\right)\mathrel{\left|\vphantom{{\psi }_i H_1\left(t\right) {\psi }_j}\right.\kern-\nulldelimiterspace}{\psi }_j\right\rangle c_j\left(0\right)e^{\frac{i}{\hbar }\left(E_i-E_j\right)t}}.
\end{equation}
The above can be readily solved once the initial conditions are specified, and subsequently the spin polarization $\mathbf{S}(t)=\langle\chi(t)|\boldsymbol{\sigma}|\chi(t)\rangle$ can be evaluated.

We first consider the system of electrons, which are initially in an admixture of the two eigen-states. The initial spin polarization vector is $\pm {\bf S}^{(0)}=(\mp\sin\gamma, \pm\cos\gamma, 0)$, corresponding to the two eigen-branches. To the first order in $H_1$, the spin polarization vector at a subsequent time $t$ is easily found to be
\begin{equation}\label{eq21}
{\bf S}(t)=\pm [{\bf S}^{(0)}+{\bf S}^{(1)}+{\bf S}^{(2)}],
\end{equation}
where
\begin{eqnarray}
{\bf S}^{(1)}= -\frac{e}{\Omega} ({\bf E}&.&\nabla_{\bf p}\gamma){\hat z},\label{eq21b}\\
{\bf S}^{(2)}(t)=\frac{2p}{\hbar } \sin{(\gamma-\theta)}&&\int^t_0 dt'{\alpha }_1(t') {\bf R}(t-t'),\label{eq21c}
\end{eqnarray}
with ${\bf R}(t)=(-\cos\gamma \sin{\Omega t}, \sin\gamma \sin{\Omega t}, \cos{\Omega t})$ being the response functions and $\Omega= 2\Delta E/\hbar$. The two terms ${\bf S}^{(1)}$ and ${\bf S}^{(2)}$ are the contributions to the spin polarization due to the applied electric field and time modulation of the Rashba coupling, respectively. To the first order in both ${\bf E}$ and $\alpha_1$, these two factors can be considered as independent driving forces which induce the spin polarization.  We focus on $\mathbf{S}^{(2)}$, the spin polarization arising from the time-modulation in the Rashba SOC. It is interesting to point out that in the absence of Dresselhauss SOC (i.e., $\beta=0$), the phase factor is simply $\gamma=\theta$, and therefore, from Eq. \eqref{eq21c}, the spin polarization $\mathbf{S}^{(2)}$ becomes zero. Hence, the time-modulation of the Rashba SOC will only induce a spin polarization if both types of SOCs, i.e., Rashba and Dresselhauss, are present in the system. In addition, we find that $\mathbf{S}^{(2)}$ is an odd function of the momentum vector $\mathbf k$, so that summing over all occupied states in momentum space would yield a zero net $z$-spin polarization. However, we will show below that the spin current may still be finite.
\begin{figure}
 \includegraphics[width=0.45\textwidth]{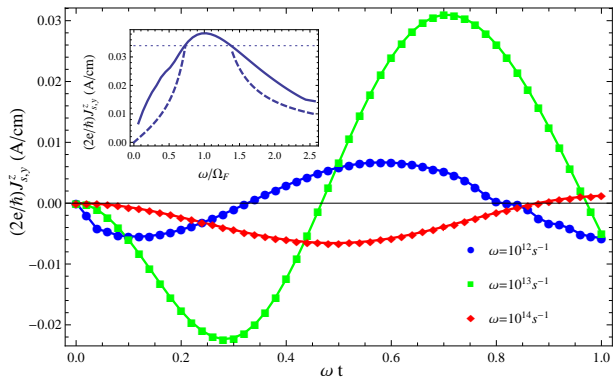}
 \caption{ Spin current generated in the system with Rashba coupling $\alpha=\alpha_0+\alpha_1\sin\omega t$. The inset shows the dependence of the spin current amplitude on the modulation frequency $\omega$ obtained numerically (solid). For comparison, we plot the analytical results for the maximum amplitude given by Eq. \eqref{eq23} (horizontal dashed) and the off-resonant amplitude given by Eq. \eqref{eq24} (bold dashed). The maximum amplitude occurs at $\omega\approx\Omega_F$.}\label{Fig2}
\end{figure}

 As discussed in previous papers,\cite{Sino04,Shen04} the modification to the electron spin due to an applied electric field can give rise to a spin Hall current (even if the net spin polarization is zero). This can be readily seen by evaluating the expression $J^z_{s,i}=(\hbar/2) (p_i/m) S^{(1)}_z$, which would recover the result of Eq. \eqref{eq15}. We apply the same expression to evaluate the spin current arising from the spin polarization $\mathbf S^{(2)}$ arising from the time-modulation of the Rashba SOC, i.e., $J^z_{s,i}=(\hbar/2) (p_i/m) S^{(2)}_z$, and average the result over the momentum space to yield the net spin current. Fig. \ref{Fig2} illustrates the ac spin current generated in the system when the Rashba coupling is modulated by a sinusoidal gate voltage, i.e. $\alpha=\alpha_0+\alpha_1 \sin\omega t$, assuming typical parameter values: $\hbar\alpha=\hbar\beta=10^{-11}\text{eV.m}$, $\alpha_1=0.1 \alpha_0$, $E_F=0.1 \text{eV}$, $m=0.05m_e$. Note that when the gate voltage is changed, the Fermi energy and effective mass will also be modified as a  consequence \cite{Nit97}. However, for simplicity, we assume a small change in the gate voltage, so that the Fermi energy and effective mass can be treated as constant. We see that the amplitude of the spin current is maximum around the resonant region $\omega\approx\Omega_F$, where $\Omega_F=2k_F\sqrt{\alpha_0^2+\beta^2}$ the Larmor frequency of electron near the Fermi surface (see inset of Fig. \ref{Fig2}).  For the special case of $\alpha_0=\beta$, this maximum amplitude can be estimated as
\begin{equation}\label{eq23}
J^z_{s,i}(\Omega_F)\sim \frac{a_1}{\Omega_F} \frac{4mE_F^2}{3\pi \hbar^2} \sqrt{\frac{E_{so}}{E_F}}=\frac{a_1 p_F^2}{6\pi\hbar},
\end{equation}
while the off-resonant amplitude is given by
\begin{equation}\label{eq24}
J^z_{s,i}(\omega)\sim \frac{a_1\omega}{\omega^2-\Omega_F^2} \frac{8mE_F^2}{3\pi^2 \hbar^2} \sqrt{\frac{E_{so}}{E_F}}.
\end{equation}
In the above, we have assumed that $E_{so}/E_F\ll 1$, where $E_{so}=m \alpha^2$ being the energy scale associated with the spin-orbit couplings, $E_F=p_F^2/2m$ the Fermi energy. The analytical results of Eqs. \eqref{eq23} and \eqref{eq24} are in approximate agreement with the numerical results (see inset of Fig. \ref{Fig2}).

\section{CONCLUSION}
In this paper, we have applied the spin force picture to a general time-dependent SOC system. By considering the Lorentz force arising from the non-Abelian gauge field of the SOC effects we establish the relation between the spin current and spin force in the system. Previously the spin force in a Rashba system was invoked to explain the Zitterbewegung motion of electrons, but no link was made to the spin Hall effect. We applied our gauge field method to show that the \emph{longitudinal} component of the spin force induces the spin Hall effect. This not only provides a physical picture of the underlying origin of the spin Hall effect, but also recovers the universal spin Hall conductivity in a constant (time invariant) Rashba system. We also showed that the spin current vanishes under strong impurity scattering, a result consistent with previous findings. However, under time modulation of the Rashba SOC a sustained spin current is obtained, the amplitude of which is proportional to the in-plane spin density and the modulation frequency. Finally, we evaluated the ac spin current generated by a Rashba SOC with a sinusoidal time variation in the absence of any dc electric field. For the special case of equal Rashba and Dresselhaus coupling, we derived the analytical expression for the magnitude of the spin current, which approximately agrees with the numerical results.\\

\begin{acknowledgments}
We gratefully acknowledge the  SERC Grant No. 092 101 0060 (R-398ְ00ְ61ֳ05) for financial support.
\end{acknowledgments}


\begin{thebibliography}{21}%
\makeatletter
\providecommand \@ifxundefined [1]{%
 \@ifx{#1\undefined}
}%
\providecommand \@ifnum [1]{%
 \ifnum #1\expandafter \@firstoftwo
 \else \expandafter \@secondoftwo
 \fi
}%
\providecommand \@ifx [1]{%
 \ifx #1\expandafter \@firstoftwo
 \else \expandafter \@secondoftwo
 \fi
}%
\providecommand \natexlab [1]{#1}%
\providecommand \enquote  [1]{``#1''}%
\providecommand \bibnamefont  [1]{#1}%
\providecommand \bibfnamefont [1]{#1}%
\providecommand \citenamefont [1]{#1}%
\providecommand \href@noop [0]{\@secondoftwo}%
\providecommand \href [0]{\begingroup \@sanitize@url \@href}%
\providecommand \@href[1]{\@@startlink{#1}\@@href}%
\providecommand \@@href[1]{\endgroup#1\@@endlink}%
\providecommand \@sanitize@url [0]{\catcode `\\12\catcode `\$12\catcode
  `\&12\catcode `\#12\catcode `\^12\catcode `\_12\catcode `\%12\relax}%
\providecommand \@@startlink[1]{}%
\providecommand \@@endlink[0]{}%
\providecommand \url  [0]{\begingroup\@sanitize@url \@url }%
\providecommand \@url [1]{\endgroup\@href {#1}{\urlprefix }}%
\providecommand \urlprefix  [0]{URL }%
\providecommand \Eprint [0]{\href }%
\providecommand \doibase [0]{http://dx.doi.org/}%
\providecommand \selectlanguage [0]{\@gobble}%
\providecommand \bibinfo  [0]{\@secondoftwo}%
\providecommand \bibfield  [0]{\@secondoftwo}%
\providecommand \translation [1]{[#1]}%
\providecommand \BibitemOpen [0]{}%
\providecommand \bibitemStop [0]{}%
\providecommand \bibitemNoStop [0]{.\EOS\space}%
\providecommand \EOS [0]{\spacefactor3000\relax}%
\providecommand \BibitemShut  [1]{\csname bibitem#1\endcsname}%
\let\auto@bib@innerbib\@empty
\bibitem [{\citenamefont {Shen}(2005)}]{Shen05}%
  \BibitemOpen
  \bibfield  {author} {\bibinfo {author} {\bibfnamefont {S.~Q.}\ \bibnamefont
  {Shen}},\ }\href@noop {} {\bibfield  {journal} {\bibinfo  {journal} {Phys.
  Rev. Lett.}\ }\textbf {\bibinfo {volume} {95}},\ \bibinfo {pages} {187203}
  (\bibinfo {year} {2005})}\BibitemShut {NoStop}%
\bibitem [{\citenamefont {Fujita}\ \emph {et~al.}(2010)\citenamefont {Fujita},
  \citenamefont {Jalil}, ,\ and\ \citenamefont {Tan}}]{Fu10}%
  \BibitemOpen
  \bibfield  {author} {\bibinfo {author} {\bibfnamefont {T.}~\bibnamefont
  {Fujita}}, \bibinfo {author} {\bibfnamefont {M.~B.~A.}\ \bibnamefont
  {Jalil}}, , \ and\ \bibinfo {author} {\bibfnamefont {S.~G.}\ \bibnamefont
  {Tan}},\ }\href@noop {} {\bibfield  {journal} {\bibinfo  {journal} {New J.
  Phys.}\ }\textbf {\bibinfo {volume} {12}},\ \bibinfo {pages} {013016}
  (\bibinfo {year} {2010})}\BibitemShut {NoStop}%
\bibitem [{\citenamefont {Aharonov}\ and\ \citenamefont {Stern}(1992)}]{Ah92}%
  \BibitemOpen
  \bibfield  {author} {\bibinfo {author} {\bibfnamefont {Y.}~\bibnamefont
  {Aharonov}}\ and\ \bibinfo {author} {\bibfnamefont {A.}~\bibnamefont
  {Stern}},\ }\href@noop {} {\bibfield  {journal} {\bibinfo  {journal} {Phys.
  Rev. Lett.}\ }\textbf {\bibinfo {volume} {69}},\ \bibinfo {pages} {3593}
  (\bibinfo {year} {1992})}\BibitemShut {NoStop}%
\bibitem [{\citenamefont {Tan}\ \emph {et~al.}(2008)\citenamefont {Tan},
  \citenamefont {Jalil}, \citenamefont {Liu},\ and\ \citenamefont
  {Fujita}}]{Tan08}%
  \BibitemOpen
  \bibfield  {author} {\bibinfo {author} {\bibfnamefont {S.~G.}\ \bibnamefont
  {Tan}}, \bibinfo {author} {\bibfnamefont {M.~B.~A.}\ \bibnamefont {Jalil}},
  \bibinfo {author} {\bibfnamefont {X.-J.}\ \bibnamefont {Liu}}, \ and\
  \bibinfo {author} {\bibfnamefont {T.}~\bibnamefont {Fujita}},\ }\href@noop {}
  {\bibfield  {journal} {\bibinfo  {journal} {Phys. Rev. B}\ }\textbf {\bibinfo
  {volume} {78}},\ \bibinfo {pages} {245321} (\bibinfo {year}
  {2008})}\BibitemShut {NoStop}%
\bibitem [{\citenamefont {Sun}\ \emph {et~al.}(2003)\citenamefont {Sun},
  \citenamefont {Guo},\ and\ \citenamefont {Wang}}]{Sun03}%
  \BibitemOpen
  \bibfield  {author} {\bibinfo {author} {\bibfnamefont {Q.~F.}\ \bibnamefont
  {Sun}}, \bibinfo {author} {\bibfnamefont {H.}~\bibnamefont {Guo}}, \ and\
  \bibinfo {author} {\bibfnamefont {J.}~\bibnamefont {Wang}},\ }\href@noop {}
  {\bibfield  {journal} {\bibinfo  {journal} {Phys. Rev. Lett.}\ }\textbf
  {\bibinfo {volume} {90}},\ \bibinfo {pages} {258301} (\bibinfo {year}
  {2003})}\BibitemShut {NoStop}%
\bibitem [{\citenamefont {Sherman}\ \emph {et~al.}(2005)\citenamefont
  {Sherman}, \citenamefont {Najmaie},\ and\ \citenamefont {Sipe}}]{Sherman05}%
  \BibitemOpen
  \bibfield  {author} {\bibinfo {author} {\bibfnamefont {E.~Y.}\ \bibnamefont
  {Sherman}}, \bibinfo {author} {\bibfnamefont {A.}~\bibnamefont {Najmaie}}, \
  and\ \bibinfo {author} {\bibfnamefont {J.~E.}\ \bibnamefont {Sipe}},\
  }\href@noop {} {\bibfield  {journal} {\bibinfo  {journal} {Appl. Phys.
  Lett.}\ }\textbf {\bibinfo {volume} {86}},\ \bibinfo {pages} {122103}
  (\bibinfo {year} {2005})}\BibitemShut {NoStop}%
\bibitem [{\citenamefont {Mal'shukov}\ \emph {et~al.}(2003)\citenamefont
  {Mal'shukov}, \citenamefont {Tang}, \citenamefont {Chu}, ,\ and\
  \citenamefont {Chao}}]{Mal03}%
  \BibitemOpen
  \bibfield  {author} {\bibinfo {author} {\bibfnamefont {A.~G.}\ \bibnamefont
  {Mal'shukov}}, \bibinfo {author} {\bibfnamefont {C.~S.}\ \bibnamefont
  {Tang}}, \bibinfo {author} {\bibfnamefont {C.~S.}\ \bibnamefont {Chu}}, , \
  and\ \bibinfo {author} {\bibfnamefont {K.~A.}\ \bibnamefont {Chao}},\
  }\href@noop {} {\bibfield  {journal} {\bibinfo  {journal} {Phys. Rev. B}\
  }\textbf {\bibinfo {volume} {68}},\ \bibinfo {pages} {233307} (\bibinfo
  {year} {2003})}\BibitemShut {NoStop}%
\bibitem [{\citenamefont {Tang}\ \emph {et~al.}(2005)\citenamefont {Tang},
  \citenamefont {Mal'shukov},\ and\ \citenamefont {Chao}}]{Tang05}%
  \BibitemOpen
  \bibfield  {author} {\bibinfo {author} {\bibfnamefont {C.~S.}\ \bibnamefont
  {Tang}}, \bibinfo {author} {\bibfnamefont {A.~G.}\ \bibnamefont
  {Mal'shukov}}, \ and\ \bibinfo {author} {\bibfnamefont {K.~A.}\ \bibnamefont
  {Chao}},\ }\href@noop {} {\bibfield  {journal} {\bibinfo  {journal} {Phys.
  Rev. B}\ }\textbf {\bibinfo {volume} {71}},\ \bibinfo {pages} {195314}
  (\bibinfo {year} {2005})}\BibitemShut {NoStop}%
\bibitem [{\citenamefont {Liang}\ \emph {et~al.}(2009)\citenamefont {Liang},
  \citenamefont {Yang},\ and\ \citenamefont {Wang}}]{Liang09}%
  \BibitemOpen
  \bibfield  {author} {\bibinfo {author} {\bibfnamefont {F.}~\bibnamefont
  {Liang}}, \bibinfo {author} {\bibfnamefont {Y.~H.}\ \bibnamefont {Yang}}, \
  and\ \bibinfo {author} {\bibfnamefont {J.}~\bibnamefont {Wang}},\ }\href@noop
  {} {\bibfield  {journal} {\bibinfo  {journal} {Eur. Phys. J. B}\ }\textbf
  {\bibinfo {volume} {69}},\ \bibinfo {pages} {337} (\bibinfo {year}
  {2009})}\BibitemShut {NoStop}%
\bibitem [{\citenamefont {Zhang}\ and\ \citenamefont {Zhu}(2013)}]{Zhang13}%
  \BibitemOpen
  \bibfield  {author} {\bibinfo {author} {\bibfnamefont {S.}~\bibnamefont
  {Zhang}}\ and\ \bibinfo {author} {\bibfnamefont {W.}~\bibnamefont {Zhu}},\
  }\href@noop {} {\bibfield  {journal} {\bibinfo  {journal} {J Phys-Condens
  Mat}\ }\textbf {\bibinfo {volume} {25}},\ \bibinfo {pages} {075302} (\bibinfo
  {year} {2013})}\BibitemShut {NoStop}%
\bibitem [{\citenamefont {Datta}\ and\ \citenamefont {Das}(1990)}]{Datta90}%
  \BibitemOpen
  \bibfield  {author} {\bibinfo {author} {\bibfnamefont {S.}~\bibnamefont
  {Datta}}\ and\ \bibinfo {author} {\bibfnamefont {B.}~\bibnamefont {Das}},\
  }\href@noop {} {\bibfield  {journal} {\bibinfo  {journal} {Appl. Phys.
  Lett.}\ }\textbf {\bibinfo {volume} {56}},\ \bibinfo {pages} {665} (\bibinfo
  {year} {1990})}\BibitemShut {NoStop}%
\bibitem [{\citenamefont {Nitta}\ \emph {et~al.}(1997)\citenamefont {Nitta},
  \citenamefont {Akazaki}, \citenamefont {Takayanagi},\ and\ \citenamefont
  {Enoki}}]{Nit97}%
  \BibitemOpen
  \bibfield  {author} {\bibinfo {author} {\bibfnamefont {J.}~\bibnamefont
  {Nitta}}, \bibinfo {author} {\bibfnamefont {T.}~\bibnamefont {Akazaki}},
  \bibinfo {author} {\bibfnamefont {H.}~\bibnamefont {Takayanagi}}, \ and\
  \bibinfo {author} {\bibfnamefont {T.}~\bibnamefont {Enoki}},\ }\href@noop {}
  {\bibfield  {journal} {\bibinfo  {journal} {Phys. Rev. Lett.}\ }\textbf
  {\bibinfo {volume} {78}},\ \bibinfo {pages} {1335} (\bibinfo {year}
  {1997})}\BibitemShut {NoStop}%
\bibitem [{\citenamefont {Grundler}(2000)}]{Gru00}%
  \BibitemOpen
  \bibfield  {author} {\bibinfo {author} {\bibfnamefont {D.}~\bibnamefont
  {Grundler}},\ }\href@noop {} {\bibfield  {journal} {\bibinfo  {journal}
  {Phys. Rev. Lett.}\ }\textbf {\bibinfo {volume} {84}},\ \bibinfo {pages}
  {6074} (\bibinfo {year} {2000})}\BibitemShut {NoStop}%
\bibitem [{\citenamefont {Sinova}\ \emph {et~al.}(2004)\citenamefont {Sinova},
  \citenamefont {Culcer}, \citenamefont {Niu}, \citenamefont {Sinitsyn},
  \citenamefont {Jungwirth},\ and\ \citenamefont {MacDonald}}]{Sino04}%
  \BibitemOpen
  \bibfield  {author} {\bibinfo {author} {\bibfnamefont {J.}~\bibnamefont
  {Sinova}}, \bibinfo {author} {\bibfnamefont {D.}~\bibnamefont {Culcer}},
  \bibinfo {author} {\bibfnamefont {Q.}~\bibnamefont {Niu}}, \bibinfo {author}
  {\bibfnamefont {N.~A.}\ \bibnamefont {Sinitsyn}}, \bibinfo {author}
  {\bibfnamefont {T.}~\bibnamefont {Jungwirth}}, \ and\ \bibinfo {author}
  {\bibfnamefont {A.~H.}\ \bibnamefont {MacDonald}},\ }\href@noop {} {\bibfield
   {journal} {\bibinfo  {journal} {Phys. Rev. Lett}\ }\textbf {\bibinfo
  {volume} {92}},\ \bibinfo {pages} {126603} (\bibinfo {year}
  {2004})}\BibitemShut {NoStop}%
\bibitem [{\citenamefont {Shen}(2004)}]{Shen04}%
  \BibitemOpen
  \bibfield  {author} {\bibinfo {author} {\bibfnamefont {S.~Q.}\ \bibnamefont
  {Shen}},\ }\href@noop {} {\bibfield  {journal} {\bibinfo  {journal} {Phys.
  Rev. B}\ }\textbf {\bibinfo {volume} {70}},\ \bibinfo {pages} {081311}
  (\bibinfo {year} {2004})}\BibitemShut {NoStop}%
\bibitem [{\citenamefont {Inoue}\ \emph {et~al.}(2004)\citenamefont {Inoue},
  \citenamefont {Bauer},\ and\ \citenamefont {Molenkamp}}]{Inoue04}%
  \BibitemOpen
  \bibfield  {author} {\bibinfo {author} {\bibfnamefont {J.~I.}\ \bibnamefont
  {Inoue}}, \bibinfo {author} {\bibfnamefont {G.~E.~W.}\ \bibnamefont {Bauer}},
  \ and\ \bibinfo {author} {\bibfnamefont {L.~W.}\ \bibnamefont {Molenkamp}},\
  }\href@noop {} {\bibfield  {journal} {\bibinfo  {journal} {Phys. Rev. B.}\
  }\textbf {\bibinfo {volume} {70}},\ \bibinfo {pages} {041303} (\bibinfo
  {year} {2004})}\BibitemShut {NoStop}%
\bibitem [{\citenamefont {Adagideli}\ and\ \citenamefont
  {Bauer}(2005)}]{Ada05}%
  \BibitemOpen
  \bibfield  {author} {\bibinfo {author} {\bibfnamefont {I.}~\bibnamefont
  {Adagideli}}\ and\ \bibinfo {author} {\bibfnamefont {G.~E.~W.}\ \bibnamefont
  {Bauer}},\ }\href@noop {} {\bibfield  {journal} {\bibinfo  {journal} {Phys.
  Rev. Lett.}\ }\textbf {\bibinfo {volume} {95}},\ \bibinfo {pages} {256602}
  (\bibinfo {year} {2005})}\BibitemShut {NoStop}%
\bibitem [{\citenamefont {Bychkov}\ and\ \citenamefont
  {Rashba}(1984)}]{Rashba84}%
  \BibitemOpen
  \bibfield  {author} {\bibinfo {author} {\bibfnamefont {Y.~A.}\ \bibnamefont
  {Bychkov}}\ and\ \bibinfo {author} {\bibfnamefont {E.~I.}\ \bibnamefont
  {Rashba}},\ }\href@noop {} {\bibfield  {journal} {\bibinfo  {journal} {J.
  Phys. C}\ }\textbf {\bibinfo {volume} {17}},\ \bibinfo {pages} {6039}
  (\bibinfo {year} {1984})}\BibitemShut {NoStop}%
\bibitem [{\citenamefont {Lommer}\ \emph {et~al.}(1988)\citenamefont {Lommer},
  \citenamefont {Malcher},\ and\ \citenamefont {Rossler}}]{Lommer88}%
  \BibitemOpen
  \bibfield  {author} {\bibinfo {author} {\bibfnamefont {G.}~\bibnamefont
  {Lommer}}, \bibinfo {author} {\bibfnamefont {F.}~\bibnamefont {Malcher}}, \
  and\ \bibinfo {author} {\bibfnamefont {U.}~\bibnamefont {Rossler}},\
  }\href@noop {} {\bibfield  {journal} {\bibinfo  {journal} {Phys. Rev. Lett.}\
  }\textbf {\bibinfo {volume} {60}},\ \bibinfo {pages} {728} (\bibinfo {year}
  {1988})}\BibitemShut {NoStop}%
\bibitem [{\citenamefont {Biswas}\ and\ \citenamefont
  {Ghosh}(2012)}]{Biswas12}%
  \BibitemOpen
  \bibfield  {author} {\bibinfo {author} {\bibfnamefont {T.}~\bibnamefont
  {Biswas}}\ and\ \bibinfo {author} {\bibfnamefont {T.~K.}\ \bibnamefont
  {Ghosh}},\ }\href@noop {} {\bibfield  {journal} {\bibinfo  {journal} {J.
  Phys.: Condens. Matter}\ }\textbf {\bibinfo {volume} {24}},\ \bibinfo {pages}
  {185304} (\bibinfo {year} {2012})}\BibitemShut {NoStop}%
\bibitem [{\citenamefont {Kohn}\ and\ \citenamefont {Luttinger}(1957)}]{KL57}%
  \BibitemOpen
  \bibfield  {author} {\bibinfo {author} {\bibfnamefont {W.}~\bibnamefont
  {Kohn}}\ and\ \bibinfo {author} {\bibfnamefont {J.~M.}\ \bibnamefont
  {Luttinger}},\ }\href@noop {} {\bibfield  {journal} {\bibinfo  {journal}
  {Phys. Rev.}\ }\textbf {\bibinfo {volume} {108}},\ \bibinfo {pages} {590}
  (\bibinfo {year} {1957})}\BibitemShut {NoStop}%
\end{thebibliography}

\providecommand{\noopsort}[1]{}\providecommand{\singleletter}[1]{#1}%

\end{document}